\documentclass[prd,preprintnumbers,twocolumn,nofootinbib,showpacs,amsmath,amssymb,floatfix]{revtex4}
\usepackage{graphicx}
\usepackage{amsfonts} 
\usepackage{amssymb}
\usepackage{bbm}
\usepackage{epsfig}
\usepackage{graphics}
\usepackage{graphicx}
\usepackage{subfigure}
\usepackage{colordvi}
\usepackage{color}

\begin{document}

\newcommand{\be}{\begin{equation}}
\newcommand{\ee}{\end{equation}}
\newcommand{\bea}{\begin{eqnarray}}
\newcommand{\eea}{\end{eqnarray}}
\newcommand{\kp}{\kappa}
\newcommand{\Om}{\Omega}
\newcommand{\de}{\Delta}
\newcommand{\eps}{\epsilon}
\newcommand{\eff}{\mathrm{eff}}
\newcommand{\nr}{\mathrm{NR}}
\newcommand{\e}{\mathrm{end}}
\newcommand{\h}{\mathcal{H}}
\newcommand{\eV}{\mathrm{eV}}
\newcommand{\JL}{\Red}
\newcommand{\ML}{\Green}
\newcommand{\SP}{\Magenta}
\newcommand{\UF}{\Blue}

\title{Model independent constraints on mass-varying neutrino scenarios}

\author{Urbano Fran\c{c}a$^1$, Massimiliano Lattanzi$^2$, Julien Lesgourgues$^3$ and Sergio Pastor$^1$}

\affiliation{$^1$Instituto de F\'{\i}sica Corpuscular  (CSIC-Universitat de Val\`{e}ncia),
Ed.\ Institutos de Investigaci\'{o}n, Apdo.\ 22085, 46071 Valencia, Spain\\
$^2$ICRA $\&$ Dip. di Fisica, Universit\'{a} di Roma "La Sapienza", P.le A. Moro 2, 00185 Roma, Italy\\
$^3$CERN, Theory Division, CH-1211 Geneva 23, Switzerland, and \\
Institut de Th\'eorie des Ph\'enom\`enes Physiques, EPFL, CH-1015 Lausanne, Switzerland, and \\
LAPTH (CNRS-Universi\'{e} de Savoie), B.P. 110, F-74941 Annecy-le-Vieux Cedex, France}


\date{\today}

\begin{abstract}
Models of dark energy in which neutrinos interact with the scalar
field supposed to be responsible for the acceleration of the universe
usually imply a variation of the neutrino masses on cosmological time
scales. In this work we propose a parameterization for the neutrino
mass variation that captures the essentials of those scenarios and
allows to constrain them in a model independent way, that is, without
resorting to any particular scalar field model.  Using WMAP 5yr data
combined with the matter power spectrum of SDSS and 2dFGRS, the limit
on the present value of the neutrino mass is $m_0 \equiv
  m_{\nu}(z=0) < 0.43 \ (0.28)$ eV at $95\%$ C.L. for the case in
  which the neutrino mass was lighter (heavier) in the past, a result
  competitive with the ones imposed for standard ({\it i.e.}, constant
  mass) neutrinos.  Moreover, for the ratio of the mass variation of
  the neutrino mass $\Delta m_{\nu}$ over the current mass $m_0$ we
  found that $\log[|\Delta m_{\nu}|/m_0] < -1.3 \ (-2.7)$ at $95\%$
  C.L. for $\Delta m_{\nu} < 0 \ (\Delta m_{\nu} > 0)$, totally
consistent with no mass variation. These stringent bounds
  on the mass variation are not related to the neutrino free-streaming
  history which may affect the matter power spectrum on small
  scales. On the contrary, they are imposed by the fact that any
  significant transfer of energy between the neutrino and dark energy
  components would lead to an instability contradicting CMB and large
  scale structure data on the largest observable scales.
\end{abstract}
                        
\preprint{IFIC/08-20, LAPTH-1246/08, CERN-PH-TH/2009-144}
\maketitle

\vspace{0.5cm}

\section{Introduction}

Since the accelerated expansion of the universe was first observed with
Type Ia supernovae (SN) \cite{Riess:1998cb,Perlmutter:1998np}, 
the case for a cosmological constant-like fluid that dominates the energy density of the 
universe has become stronger and is well established by now with the new pieces of data gathered
\cite{Frieman:2008sn}.

Several candidates for the accelerating component of the universe,
generically dubbed dark energy (DE), have been proposed
\cite{Frieman:2008sn,DEreview1,DEreview2,Caldwell:2009ix}, but
understanding them theoretically and observationally has proven to be
challenging.  On the theoretical side, explaining the small value of
the observed dark energy density component, $\rho_{\phi} \sim
(10^{-3}$ eV$)^4$, as well as the fact that both dark energy and
matter densities contribute significantly to the energy budget of the
present universe requires in general a strong fine tuning on the
overall scale of the dark energy models. In the case in which the dark
energy is assumed to be a scalar field $\phi$ slowly rolling down its
flat potential $V(\phi)$, the so-called quintessence models
\cite{Caldwell:1997ii}, the effective mass of the field has to be
taken of the order $m_{\phi}= | d^2V(\phi)/d\phi^2|^{1/2} \sim
10^{-33}$ eV for fields with vacuum expectation values of the order
of the Planck mass.

On the observational side, choosing among the dark energy models is a
complicated task \cite{Linder:2008pp}. Most of them can mimic a
cosmological constant  at late times (that is, an equation of state
$w_{\phi}\equiv p_{\phi}/\rho_{\phi} = -1$)
\cite{Albrecht:2006um}, and all data until now are perfectly
consistent with this limit.  In this sense,
looking for different imprints that could favor the existence of a
particular model of dark energy is a path worth taking.  

Our goal in this paper consists in understanding whether the 
so-called Mass-Varying Neutrinos
(MaVaNs) scenario
\cite{Gu2003,Fardon:2003eh,Peccei:2004sz,Amendola:2007yx,Wetterich:2007kr}
could be constrained not only via the dark energy effects, but also by
indirect signs of the neutrino mass variation during cosmological
evolution, since neutrinos play a key role in several epochs
\cite{Hannestad:2006zg,Lesgourgues:2006nd}.  An
indication of the variation of the neutrino mass would certainly tend
to favor this models (at least on a theoretical basis) with respect to
most DE models. One should keep in mind that MaVaNs scenarios can suffer from stability
issues for the neutrino perturbations \cite{Afshordi:2005ym}, although
there is a wide class of models and couplings that avoid this problem
\cite{Bjaelde:2007ki, Bean:2007nx, Bean:2007ny, Bean:2008ac, Bernardini:2008pn}.

Similar analyses have been made in the past, but they have either
assumed particular models for the interaction between the neutrinos
and the DE field \cite{Brookfield:2005td,Brookfield:2005bz,Ichiki:2007ng}, or chosen a parameterization that
does not reflect the richness of the possible behavior of the neutrino
mass variations \cite{Zhao:2006zf}.

In order to be able to deal with a large number of models, instead of
focusing on a particular model for the coupling between the DE field
and the neutrino sector, we choose to parameterize the neutrino mass
variation to place general and robust constraints on the
MaVaNs scenario. In this sense, our work complements previous analyses
by assuming a realistic and generic parameterization for the neutrino
mass, designed in such a way to probe almost all the different regimes and models
within the same framework. In particular, our parameterization allows
for fast and slow mass transitions between two values of the neutrino
mass, and it takes into account that the neutrino mass variation should
start when the coupled neutrinos change their behavior from
relativistic to nonrelativistic species. We  can mimic
different neutrino-dark energy couplings and  allow for  
almost any monotonic behavior in the neutrino mass, 
placing reliable constraints on this scenario in a model independent way.

Our work is organized as follows: in Section \ref{sec:mvn} we give a
brief review of the MaVaNs scenario and its main equations.  In Section
\ref{sec:param} we present our parameterization with the results for the
background and the perturbation equations obtained within this
context. The results of our comparison of the numerical results
with the data and the discussion of its main implications are shown
in section \ref{sec:results}.  Finally, in section \ref{sec:conc} the
main conclusions and possible future directions are discussed.


\section{Mass-varying neutrinos} \label{sec:mvn}

In what follows, we consider a homogeneous
and isotropic universe with a Robertson-Walker flat metric, $ds^2 = a^2 
\left( d\tau^2 + dr^2 + r^2 d\Omega^2 \right)$, where  $\tau$ is
the conformal time, that can be written in terms of the cosmic
time $t$ and scale factor $a$ as $d\tau= dt/a$, in  natural
units ($\hbar=c=k_B=1$). In this
case, the Friedmann equations read
\bea \label{eq:friedmann}
\mathcal{H}^2 &=& \left( \frac{\dot{a}}{a}\right)^2= \frac{a^2}{3m_p^2}\rho,\\
\dot{\mathcal{H}}&=&-\frac{a^2}{6m_p^2} \left(\rho+3p \right),
\eea
where the dot denotes a derivative with respect to conformal time, and
the reduced Planck mass is $m_p = 1/ \sqrt{8 \pi G} = 2.436 \times
10^{18}$ GeV. As usual, $\rho$ and $p$ correspond to the total energy
density and pressure of the cosmic fluid, respectively. The 
neutrino mass in the models we are interested in is a function of
the scalar field $\phi$ that plays the role of the dark energy, and
can be written as 
\be \label{eq:mphi}
m_{\nu}(\phi) = M_{\nu} f(\phi) \ ,
\ee
where $M_{\nu}$ is a constant and 
different models are represented by distinct $f(\phi)$.

The fluid equation of the neutrino species can be directly obtained
from the Boltzmann equation for its distribution function \cite{Brookfield:2005bz}, 
\be \label{eq:fluidNU}
\dot{\rho}_{\nu}  + 3 \mathcal{H} \rho_{\nu} \left( 1+ w_{\nu}\right) =  \alpha(\phi) \dot{\phi} 
\left( \rho_{\nu} - 3 p_{\nu}\right)\ ,\\
\ee
where $\alpha(\phi) = d\ln [m_{\nu}(\phi)] / d\phi$ takes into account the
variation of the neutrino mass, and $w_x = p_x/\rho_x$ is the equation of state of the 
species $x$. For completeness and later use,
we will define $\Omega_{x0} = \rho_x/\rho_{c0}$, the
standard density parameter, where the current critical
density is given by $\rho_{c0}= 3 H_0^2 m_p^2 = 8.099 \ h^2 \times 10^{-11}$ eV$^4$ and
$H_0 = 100 \ h$ km s$^{-1}$ Mpc$^{-1}$ is the Hubble constant.

Since the total energy momentum tensor is conserved, the dark energy
fluid equation also presents an extra right-hand side term proportional
to the neutrino energy momentum tensor trace, 
$T_{(\nu)\alpha}^{\alpha}= \left( \rho_{\nu} - 3 p_{\nu}\right)$,  and
can be written as
\be \label{eq:fluidDE}
\dot{\rho}_{\phi}  +  3 \mathcal{H} \rho_{\phi} \left( 1 + w_{\phi}\right) = - \alpha(\phi) \dot{\phi} 
\left( \rho_{\nu} - 3 p_{\nu}\right)\ .\\
\ee
For a homogeneous and isotropic scalar field, the energy density and 
pressure are given by
\be \label{eq: sf}
\rho_{\phi}= \frac{\dot{\phi}^2}{2 a^2} + V(\phi)\ , \ \ \ \ \ 
p_{\phi}= \frac{\dot{\phi}^2}{2 a^2} - V(\phi) \ ,
\ee
and both equations lead to 
the standard cosmological Klein-Gordon equation for an interacting 
scalar field, namely,
\be \label{eq:kg}
\Ddot{\phi} + 2 \mathcal{H} \dot{\phi} + a^2 \frac{dV(\phi)}{d\phi} = 
- a^2 \alpha(\phi)\left( \rho_{\nu} - 3 p_{\nu}\right)\ .
\ee
From the above equations one sees that, given a potential $V(\phi)$ for
the scalar field and a field-dependent mass term $m_{\nu} (\phi)$ for the 
neutrino mass, the coupled system given by equations (\ref{eq:friedmann}), 
(\ref{eq:fluidNU}), and (\ref{eq:kg}), together with the fluid equations for the 
baryonic matter, cold dark matter and radiation (photons and other massless species) 
can be numerically solved \cite{Brookfield:2005bz}. Notice that a similar approach has been
used for a possible variation of
the dark matter mass \cite{carroll97} and its possible 
interaction with the dark energy \cite{amendola00,amendola01}, with
several interesting phenomenological 
ramifications \cite{farrar04,prd04,huey04,das05,Quartin:2008px,LaVacca2009}.

Following \cite{prd04,das05}, 
equations (\ref{eq:fluidNU}) and (\ref{eq:fluidDE}) can be rewritten
in the standard form,
\bea \label{eq:weff}
\dot{\rho}_{\nu} & + & 3 \mathcal{H} \rho_{\nu} \left( 1 + w^{(\eff)}_{\nu} \right) = 0 \ , \nonumber \\[-3mm]
& & \\[-3mm]
\dot{\rho}_{\phi} & + & 3 \mathcal{H} \rho_{\phi} \left( 1 + w^{(\eff)}_{\phi} \right) = 0 \ , \nonumber
\eea
if one defines the {\it effective} equation of state of neutrinos and DE as
\bea \label{eq:effphi}
w^{(\eff)}_{\nu} & = &  \frac{p_{\nu}}{\rho_{\nu}} - 
\frac{\alpha(\phi) \dot{\phi} \left( \rho_{\nu} - 3 p_{\nu}\right)}{3 \mathcal{H} \rho_{\nu}}\ ,\nonumber \\[-3mm]
& & \\[-3mm]
w^{(\eff)}_{\phi} & = & \frac{p_{\phi}}{\rho_{\phi}} + 
\frac{\alpha(\phi) \dot{\phi} \left( \rho_{\nu} - 3 p_{\nu}\right)}{3 \mathcal{H}\rho_{\phi}}\ \nonumber.
\eea
The effective equation of state can be understood in terms of the
dilution of the energy density of the species. In the standard noncoupled
case, the energy density of a fluid with a given constant equation of 
state $w$ scales as $\rho \propto a^{-3(1 +w)}$. However, in the case of
interacting fluids, one should also take into account the energy transfer between
them, and the energy density in this case will be given by
\be
\rho(z) = \rho_0 \exp\left[3 \int_0^{z} \left( 1+ w^{(\eff)}(z')\right) d \ln(1+z') \right] \ ,
\ee
where the index 0 denotes the current value of a parameter, and the
redshift $z$ is defined by the expansion of the scale factor, $a = a_0
(1+z)^{-1} $ (in the rest of this work we will assume $a_0=1$).
For a constant effective equation of state one obtains the standard
result, $\rho \propto a^{-3(1 +w^{(\eff)})}$, as expected.

Notice that this mismatch between the effective and standard DE
equations of state could be responsible for the ``phantom behavior''
suggested by supernovae data when fitting it using a cosmological model
with noninteracting components \cite{das05}. This effect could be
observable if dark energy was coupled to the dominant dark matter
component. For the models discussed here, however, it cannot be significant: the
neutrino fraction today ($\Omega_{\nu 0}/\Omega_{\phi 0} \sim
10^{-2}$) is too small to induce an ``effective phantom-like''
behavior.

As we commented before, the analysis until now dealt mainly with
particular models, that is, with particular functional forms of the
dark energy potential $V(\phi)$ and field dependence of the neutrino
mass $\alpha(\phi)$. A noticeable exception 
is the analysis of Ref.~\cite{Zhao:2006zf}, in which the authors 
use a parameterization for the neutrino mass
{\it a l\`{a}} Chevallier-Polarski-Linder (CPL)
\cite{Albrecht:2006um,Chevallier:2000qy,Linder:2002et}:  $m_{\nu}(a)= m_{\nu 0} + m_{\nu 1} (1-a)$.
However, although the CPL parameterization works well for the dark
energy equation of state, it cannot reproduce the main features of the
mass variation in the case of variable mass particle models. In the
case of the models discussed here, for instance, the 
mass variation is related to
the relativistic/nonrelativistic nature of the coupled neutrino
species. With a CPL mass parameterization, the transition from $m_1$ to $m_0$
always takes place around $z \sim 1$, which is in fact only compatible with
masses as small as $10^{-3}$ eV. Hence, the
CPL mass parameterization is not suited for a self-consistent
exploration of all interesting possibilites.

One of the goals in this paper is to propose and test a
parameterization that allows for a realistic simulation of
mass-varying scenarios in a model independent way, with the minimum
possible number of parameters, as explained in the next section.


\section{Model independent approach} \label{sec:param}

\subsection{Background equations}

As usual, the neutrino energy density and pressure are given in terms
of the zero order Fermi-Dirac distribution function by
\be
f^0(q)=\frac{g_{\nu}}{e^{q/T_{\nu 0}}+1} \ ,
\ee
where $q= a p$ denotes the modulus of the comoving
momentum $q_i = q n_i$ ($\delta^{ij}n_in_j = 1$), $g_\nu$ corresponds to the number of neutrino
degrees of freedom, and $T_{\nu 0}$ is the present neutrino background
temperature.  Notice that in the neutrino distribution function we
have used the fact that the neutrinos decouple very early in the
history of the universe while they are relativistic, and therefore
their equilibrium distribution depends on the comoving momentum, but
not on the mass \cite{Lesgourgues:2006nd}. In what follows we have
neglected the small spectral distortions arising from
non-instantaneous neutrino decoupling \cite{Mangano:2005cc}.  Thus,
the neutrino energy density and pressure are given by
\be \label{eq:density}
\rho_\nu=\frac{1}{a^4}\int \frac{dq}{(2\pi)^3}\, d\Omega \, q^2 \eps f^0(q) \ ,
\ee
\be \label{eq:pressure}
p_\nu=\frac{1}{3a^4}\int \frac{dq}{(2\pi)^3} \, d\Omega \, q^2 f^0(q) \frac{q^2}{\eps} \ ,
\ee
where $\eps^2 = q^2 + m_\nu^2(a)a^2$ (assuming that $m_{\nu}$ depends
only on the scale factor).
Taking the time-derivative of the energy density, one can then obtain the fluid equation
for the neutrinos,
\begin{equation}\label{eq:nuenergy}
\dot{\rho}_\nu  + 3 \h \left( \rho_\nu + p_\nu\right) = \frac{d \ln m_\nu(u)}{du} \h  \left( \rho_\nu  - 3 p_\nu \right) \ ,
\end{equation}
where $u \equiv \ln a = - \ln(1+z)$ is the number of e-folds
counted back from today. Due to the conservation of the total
energy momentum tensor, the dark energy fluid equation is then given
by
\begin{equation}\label{eq:deenergy}
\dot{\rho}_{\phi}\  + 3 \h \rho_{\phi} \left( 1 + w_{\phi}\right) = 
- \frac{d \ln m_\nu(u)}{du} \h  \left( \rho_\nu  - 3 p_\nu \right)\ .
\end{equation}
We can write the effective equations of state, 
defined in eqs.~(\ref{eq:weff}), as
\bea \label{eq:effeos}
w^{\eff}_\nu & = &  \frac{p_{\nu}}{\rho_{\nu}} 
- \frac{d \ln m_\nu(u)}{du}  \left( \frac{1}{3} - \frac{p_{\nu}}{\rho_{\nu}} \right) \ ,\nonumber \\ [-2mm]
& & \\[-2mm]
w^{\eff}_{\phi} & = & \frac{p_{\phi}}{\rho_{\phi}} +    
\left( \frac{\Omega_{\nu}}{\Omega_{\phi}} \right) \frac{d \ln m_\nu(u)}{d u}  \left( \frac{1}{3} - \frac{p_{\nu}}{\rho_{\nu}}\right)\ \nonumber.
\eea
The above results only assume that the neutrino mass depends on the
scale factor $a$, and up to this point, we have not chosen any
particular parameterization. Concerning the particle physics models,
it is important to notice that starting from a value of $w_{\phi}$ and
a function $m_{\nu}(a)$ one could, at least in principle, reconstruct
the scalar potential and the scalar interaction with neutrinos
following an approach similar to the one in
Ref.~\cite{Rosenfeld:2007ri}.

\subsection{Mass variation parameters} \label{subsec:mvp}

Some of the main features of the MaVaNs scenario are: {\it (i)}
that the dark energy field gets kicked and moves away from its minimum
(if $m_\phi > H$) or from its previous slow-rolling trajectory (if
$m_\phi < H$) when the neutrinos become non-relativistic, very much
like the case when it is coupled to the full matter content of the
universe in the so-called {\it chameleon} scenarios \cite{Brax:2004qh};
and {\it (ii)} 
that as a consequence, the coupling with the scalar field 
generates a neutrino mass variation at that time.
Any parameterization that intends to mimic scalar
field models interacting with a mass-varying
particle (neutrinos, in our case) for the large
redshift range to which the data is sensitive
should at least take into account those characteristics. 
Moreover, the variation of the mass in most models (see
\cite{Brookfield:2005bz}, for instance) can be well approximated by a
transition between two periods: an earlier one, in which the mass is
given by $m_1$, and the present epoch, in which the mass is given by
$m_0$ (we will not consider here models
in which the neutrino mass behavior is nonmonotonic). The transition for this 
parameterization, as mentioned before, starts when neutrinos
become nonrelativistic, which corresponds approximately to
\bea \label{eq:znr}
z_{\nr} \approx  1.40  \left(\frac{1 \ \mathrm{eV}}{3 \ T_{\gamma0}} \right)
\left( \frac{m_{1}}{1 \ \mathrm{eV}} \right)
\approx 2 \times 10^3 \left(\frac{m_{1}}{1 \ \mathrm{eV}} \right) 
\eea
where $m_1$ corresponds to the mass of the neutrino during the period
in which it is a relativistic species. Before $z_\nr$ we can treat the
neutrino mass as essentially constant, since the right-hand side (RHS)
of the fluid equation is negligible compared to the
left-hand side (LHS), and therefore there is no observable
signature of a possible mass variation.

\begin{figure}[ht]
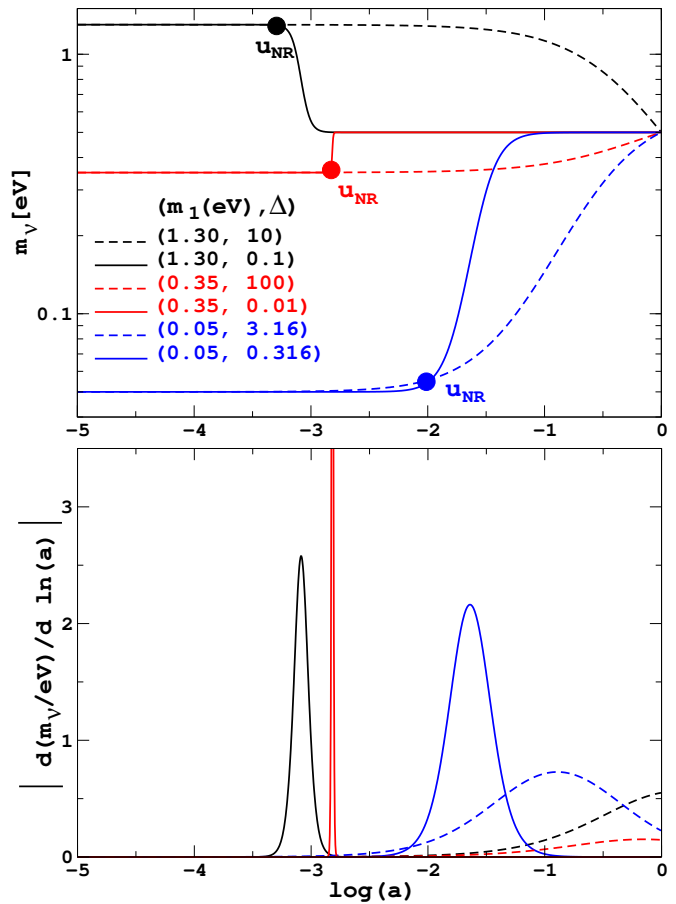

\begin{tabular}{r}
\includegraphics[scale=0.37]{fig1a.eps} \\
\includegraphics[scale=0.37]{fig1b.eps}
\end{tabular}
\caption{   \label{fig:mvn} (Color online) Neutrino mass behavior
for the parameterization given by equation (\ref{eq:paramnumass}). 
{\it Top panel}: Neutrino mass as a function of $\log(a)=u/\ln(10)$
for models with 
$m_0=0.5$ eV and different values of $m_1$ and $\Delta$. 
{\it Bottom panel:} Neutrino mass variation for the same parameters
as in the top panel.}
\end{figure}

When the neutrinos become nonrelativistic, the RHS of the DE and
neutrino fluid equations becomes important, and the neutrino mass
starts varying. In order to model this variation, we use two
parameters, namely the current neutrino mass, $m_0$, and 
$\de$, a quantity related to the amount of time that it takes to complete the
transition from $m_1$ to $m_0$. That behavior resembles very much the
parameterization of the dark energy equation of state discussed in
\cite{Corasaniti:2002vg}, except for the
fact that in our case the transition for the mass can be very slow,
taking several e-folds to complete, and must be
triggered by the time of the nonrelativistic transition, given by
equation (\ref{eq:znr}).  Defining $f=[1+e^{-[u \ (1+ \de)- u_{\nr}]/ \de}]^{-1}$ and $f_{*}=[1+ e^{u_{\nr}/ \de}]^{-1}$ we can use the form
\be \label{eq:paramnumass}
m_{\nu} = m_0 + (m_1 - m_0) \times \Gamma(u, u_{\nr},\de) \ ,
\ee
where
\bea \label{eq:Gamma}
\Gamma(u, u_{\nr},\de) & = & 1 - \frac{f}{f_{*}} \nonumber \\[-2mm] 
&& \\[-2mm]
& = & \left[ 1 - \frac{1+ e^{u_{\nr}/ \de}}
{1+e^{-[u \ (1+ \de)- u_{\nr}]/ \de} } \right] \ . \nonumber
\eea
Starting at $u_{\nr}= - \ln (1 + z_{\nr})$, the function $\Gamma(u,
u_{\nr},\de)$ decreases from 1 to 0, with a velocity that depends on
$\de$.  The top panel in Figure \ref{fig:mvn} gives the behavior of
eq.\ (\ref{eq:paramnumass}) with different parameters; the bottom
panels shows that in this parametrization, the derivative of the mass
with respect to e-fold number resembles a Gaussian function.
The peak of the quantity
$dm/du$ occurs at the value $\bar u = u_{\nr}/(1+\de)$; hence, for
$\Delta \ll 1$, the mass variation takes place immediately after the
non-relativistic transition ($\bar{u} \simeq u_{\nr}$) and lasts a
fraction of e-folds (roughly, $3\Delta$ e-folds); for $1 \leq \Delta
\leq |u_{\nr}|$ the variation is smooth and centered on some
intermediate redshift between $z_{NR}$ and 0; while for $\Delta \gg
|u_{NR}|$, the transition is still on-going today, and the present epoch
roughly coincides with the maximum variation. 

Although the functional form of $\Gamma$, eq.\ (\ref{eq:Gamma}), seems
complicated, one should note that it is one of the simplest forms
satisfying our requirements with a minimal number of parameters.
An example that could look simpler, but that for practical purposes
is not, would be to assume that the two plateaus are linked together by
a straight line.  In this case, we would need a parameterization of
the form
\begin{equation*}
m_{\nu} \! = \! \left\{ \begin{array}{llr}
\ m_1                    &\! , \!&   u  <  u_{\nr} \ ,\\[2mm]
m_0+ (m_1 -  m_0)\left[{ \frac{u - u_\e}{u_\nr - u_\e} }\right] 
 & \! , \! &   u_{\nr} \leq u \leq u_{\mathrm{end}} \ , \\[2mm]
\ m_0                   & \! , \! &      u  >  u_{\mathrm{end}}
\end {array} \right.
\end{equation*}
where $u_{\e}$ corresponds to the chosen redshift in which the 
transition stops. Notice that in this case not only we still
have three parameters to describe the mass variation, but also
the function is not smoooth. Moreover, 
the derivative of the mass with respect to $u$ gives a 
top-hat-like function which is discontinuous 
at both $u_\nr$ and $u_\e$. 
In this sense, it seemed to us that equation 
(\ref{eq:paramnumass}) would give us 
the best ``price-to-earnings ratio'' among the possibilities to
use phenomenologically motivated parameterizations for the 
mass-varying neutrinos, although certainly there could be 
similar proposals equally viable, such as for instance the possibility of adapting 
for the mass variation
the parameterization used for the dark energy equation of state in 
\cite{Douspis:2006rs, Linden:2008mf}. There, the
transition between two constant values of the equation of state exhibits  a 
$\tanh \left[ \Gamma_t (u-u_t)\right]$ dependence, where $\Gamma_t$ is responsible
for the duration of the transition and $u_t$ is related to its half-way point.

In the rest of our analysis, we will use a couple of extra assumptions
that need to be taken into account when going through our results.
First, we will consider that only one of the three neutrino species is
interacting with the dark energy field, that is, only one of the mass
eigenstates has a variable mass. The reason for this approximation is
twofold: it is a simpler case (compared to the case with 3
varying-mass neutrinos), since instead of 6 extra parameters with
respect to the case of constant mass, we have only 2, namely the early
mass of the neutrino whose mass is varying, $m_1$, and the velocity of
the transition, related to $\de$.  

Besides simplicity, the current choice 
is the only one allowed presently in the case in
which neutrinos were heavier in the past.
Indeed, we expect our
stronger constraints to come from those scenarios, especially if the
neutrino species behaves as a
nonrelativistic component at the time of radiation-matter equality,
given by $1+z_{\mathrm{eq}} \sim 4.05\times 10^4 (\Omega_{c0}h^2 +
\Omega_{b0}h^2 )/(1+0.23 N_{\mathrm{eff}})$
(here the indexes $c$ and $b$ stand for cold
dark matter and baryons, respectively, and $ N_{\mathrm{eff}}$ is the
effective number of relativistic neutrinos). 
Taking the three neutrino species to be nonrelativistic at equality would change
significantly the value of $z_{\mathrm{eq}}$, contradicting CMB data
(according to WMAP5, $1+z_{\mathrm{eq}} = 3141^{+ 154}_{-157}$ 
(68$\%$ C.L.) \cite{Komatsu:2008hk}).
Instead, a single neutrino species is still marginally 
allowed to be non-relativistic at that time.

To simplify the analysis, we also assumed that the dark energy field, when not interacting
with the neutrinos, reached already the so-called scaling solution 
(see, {\it e.g.}, \cite{DEreview1} and references therein),
i.e., the dark energy equation of state $w_{\phi}$ in eq.\ (\ref{eq:deenergy}) is constant
in the absence of interaction. Notice however that when the 
neutrinos become non-relativistic
the dark energy fluid receives the analogous of the chameleon kicks
we mentioned before, and the dark energy effective equation of state, 
eq.\ (\ref{eq:effeos}), does vary for this period in a consistent way. 

\begin{figure}[ht]
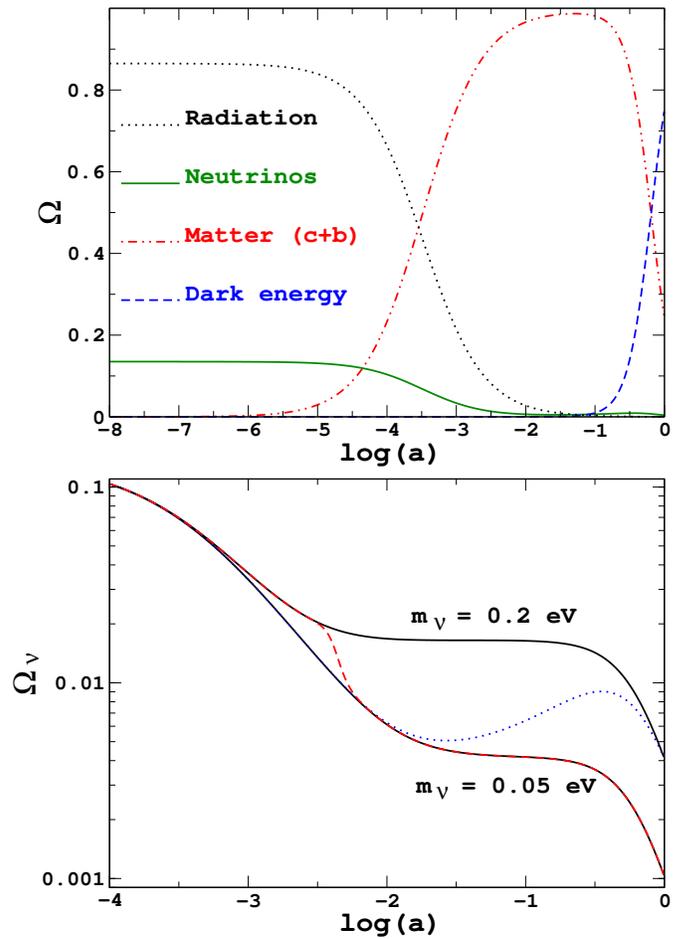

\begin{tabular}{r}
\includegraphics[scale=0.37]{fig2a.eps}\\
\includegraphics[scale=0.37]{fig2b.eps}
\end{tabular}
\caption{   \label{fig:omegas} (Color online)
{\it Top panel}: Density parameters for the different components
of the universe versus $\log(a)=u/\ln(10)$ in a model 
with  $m_1=0.05$ eV, $m_0 = 0.2$ eV, $\Delta = 10$, and all the
other parameters consistent with present data. The radiation curve include
photons and two massless neutrino species, and matter stands for cold 
dark matter and baryons. The bump in the neutrino density close to 
$\log(a)=-0.5$
is due to the increasing neutrino mass.
{\it Bottom panel:} Density parameters for two different mass-varying neutrino
models. The solid black curves show the density parameter variation
for two distinct constant mass models, with masses $m_{\nu} = 0.05$ eV and 
$m_{\nu} = 0.2$ eV.  The dashed (red) curve shows a model in which the mass varies
from $m_1= 0.2$ eV to $m_0 = 0.05$ eV, with $\Delta= 0.1$, and the dotted (blue) line
corresponds a model with $m_1= 0.05$ eV to $m_0 = 0.2$ eV, with $\Delta= 10$.
}
\end{figure}

The upper panel of Figure \ref{fig:omegas} shows  how the density parameters 
of the different components of the universe evolve in time, in a typical 
MaVaNs model. The lower panel displays a comparison between mass-varying 
and constant mass models, in particular during the transition from $m_1$ to $m_0$.
As one would expect, far from the time of the transition, the densities
evolve as they would do in the constant mass case.

\subsection{Perturbation equations}

The next step is to calculate the cosmological
perturbation equations and their evolution
using this parameterization.  We chose to work in the synchronous gauge,
and our conventions follow the ones by  Ma and
Bertschinger \cite{Ma:1995ey}. In this case, the perturbed metric is given by 
\begin{equation}
ds^2 = -a^2 d\tau^2 + a^2 \left(\delta_{ij} + h_{ij} \right) dx^i dx^j \ .
\end{equation}
In this gauge, the equation  for the three-momentum of the neutrinos reads 
\cite{Ichiki:2007ng}
\be \label{eq:eq-b}
 \frac{dq}{d\tau} = -\frac{1}{2}q \dot{h}_{ij} n_i n_j - a^2
  \frac{m_{\nu}^2}{q} \beta \frac{\partial \rho_{\phi}}{\partial x^i}\frac{\partial x^i}{d \tau} \ ,
\ee
where, as in  equation (\ref{eq:fluidNU}), we define
\be \label{eq:beta}
\beta(a) \equiv \frac{d\ln m_\nu}{d\rho_\phi}= 
\frac{d\ln m_\nu}{d \ln a} \left( \frac{d\rho_\phi}{d \ln a} \right)^{-1} \ .
\ee
Since the neutrino phase space distribution \cite{Ma:1995ey} can be
written as $f\left(x^i,q, n_j,\tau\right)= f^0(q)\left[1+\Psi
  \left(x^i,q, n_j,\tau\right)\right]$, one can show that the first
order Boltzmann equation for a massive neutrino species, after Fourier transformation, is given
by \cite{Brookfield:2005bz,Ichiki:2007ng} 
\bea \label{eq:nupert}
 \frac{\partial \Psi}{\partial
 \tau} & + & i\frac{q}{\eps}({\bf \hat{n}}\cdot{\bf k})\Psi+\left(\dot\eta-({\bf \hat{k}}
\cdot{\bf \hat{n}})^2\frac{\dot{h}+6\dot{\eta}}{2}\right)\frac{d \ln
 f^0}{d \ln q} \nonumber \\ [-2mm]
& & \\ [-2mm] 
& = & -i\beta \frac{qk}{\eps} ({\bf \hat{n}}\cdot{\bf k})\frac{a^2
 m_{\nu}^2}{q^2}\frac{d \ln
 f^0}{d \ln q} \delta\rho_{\phi} \ ,\nonumber 
\eea
where $\eta$ and $h$ are the synchronous potentials in the Fourier space.
Notice that the perturbed neutrino energy density and pressure are also going to be 
modified due to the interaction, and are written as
\bea \label{eq:dens_pres_pert}
\delta \rho_{\nu} & = & \frac{1}{a^4} 
\int \frac{d^3q}{(2\pi)^3}  f^0 
\left( \epsilon \Psi + \beta \frac{m_{\nu}^2 a^2}{\epsilon} \delta \rho_{\phi}  \right) \ , \\
3 \delta p_\nu & = & \frac{1}{a^4} \int \frac{d^3q}{(2\pi)^3}  f^0
\left(\frac{q^2}{\eps}\Psi - \beta \frac{q^ 2 m_{\nu}^2 a^2}{\epsilon^3} \delta \rho_{\phi} \right) .
\eea
This extra term comes from the fact that the comoving energy $\epsilon$ depends on the
dark energy density, leading to an extra-term which is proportional to $\beta$.

Moreover, if we expand the perturbation $\Psi \left({\bf k},q, {\bf n},\tau\right)$ in a Legendre
series \cite{Ma:1995ey}, the neutrino hierarchy equations will read,
\bea \label{eq:hierarchy}
\dot{\Psi}_0& = & -\frac{q k}{\eps}\Psi_1
+\frac{\dot h}{6}\frac{d \ln{f^0}}{d\ln{q}} \ , \nonumber \\
\dot{\Psi}_1& = & \frac{qk}{3 \eps}\left(\Psi_0-2\Psi_2\right)
+ \kappa \ , \\
\dot{\Psi}_2& = & \frac{qk}{5 \eps}(2\Psi_1-3\Psi_3)
-\left(\frac{1}{15}\dot{h}+\frac{2}{5}\dot{\eta}\right)\frac{d \ln{f^0}}{d \ln{q}} , \nonumber \\
\dot{\Psi}_\ell& = & \frac{qk}{(2\ell+1) \eps}\left[\ell\Psi_{\ell-1}
-(\ell+1)\Psi_{\ell+1} \right]\ . \nonumber
\eea
where
\be \label{eq:kappa}
\kappa = -\frac{1}{3} \beta \ \frac{q k}{\epsilon}\frac{a^2 m_{\nu}^2}{q^2}
 \frac{d \ln
 f^0}{d\ln q} \delta \rho_{\phi}\ .
\ee
%


For the dark energy, we use the ``fluid approach'' \cite{Hu:1998kj} (see also
\cite{Bean:2003fb,Hannestad:2005ak,Koivisto:2005mm}),
so that the density and velocity perturbations are given by,
\begin{widetext}
\begin{equation} \label{eq:depert1}
\dot{\delta}_{\phi} = \frac{3 {\cal H} (w_{\phi} - \hat{c}_{\phi}^2) 
\left( \delta_{\phi} 
+ \frac{3 {\cal H} (1+w_\phi)}{
{ 1 + \beta \rho_{\nu} (1 - 3 w_{\nu})}}
\frac{\theta_{\phi}}{k^2}\right)
- (1 + w_{\phi}) \left( \theta_{\phi} + \frac{\dot{h}}{2} \right)
{
- \left( \frac{\rho_{\nu}}{\rho_{\phi}} \right)
\left[ \beta \dot{\rho}_{\phi} (1-3c_\nu^2) \delta_{\nu} + \dot{\beta}
\rho_{\phi} (1 - 3 w_{\nu}) \delta_{\phi} \right]}}
{{ 1 + \beta \rho_{\nu} (1 - 3 w_{\nu})}} \ ,
\end{equation}
\begin{equation} \label{eq:depert2}
\dot{\theta}_{\phi} = -
\left[
\frac{{\cal H} (1 - 3 \hat{c}_{\phi}^2) 
{
+ \beta \rho_{\nu} (1 - 3 w_{\nu}) {\cal H} (1 - 3 w_\phi)}}
{{ 1 + \beta \rho_{\nu} (1 - 3 w_{\nu})}} 
\right]
\theta_{\phi}
+ \frac{k^2}{1 + w_{\phi}} 
\hat{c}_{\phi}^2 \delta_{\phi} 
{
- \beta (1 - 3 w_\nu) \left(\frac{\rho_\nu}{\rho_\phi} \right)
\left[ \frac{k^2}{1+w_\phi} \rho_\phi \delta_\phi
- \dot{\rho}_\phi \theta_\phi \right] \ ,}
\end{equation}
\end{widetext}
where the dark energy anisotropic stress is assumed to be zero \cite{Mota:2007sz}, and the sound speed $\hat{c}_{\phi}^2$ is defined in the frame comoving with the dark energy fluid \cite{Weller:2003hw}.
So, in the synchronous gauge, the quantity $c_{\phi}^2 \equiv \delta p_{\phi}/ \delta \rho_{\phi}$ is related to $\hat{c}_{\phi}^2$ through
\begin{equation}
c_{\phi}^2 \delta_{\phi} 
= \hat{c}_{\phi}^2 
\left(\delta_{\phi} - \frac{\dot{\rho}_{\phi}}{\rho_{\phi}} \frac{\theta_{\phi}}{k^2} \right) 
+ w_{\phi} \frac{\dot{\rho}_{\phi}}{\rho_{\phi}} \frac{\theta_{\phi}}{k^2} \ .
\end{equation}
In addition, from eqs. (\ref{eq:deenergy}) and 
(\ref{eq:beta}), we have that 
\be \label{eq:dotde}
\frac{\dot{\rho}_{\phi}}{ \rho_{\phi}}= \frac{-3 \h (1+ w_{\phi})}{1 + \beta \rho_{\nu}(1-3 w_{\nu})} \ .
\ee
%


\section{Results and Discussion} \label{sec:results}

\subsection{Numerical approach}
  
Equipped with the background and perturbation equations, we can 
study this scenario by modifying the numerical packages that
evaluate the CMB anisotropies and the matter
power spectrum. In particular, we modified the CAMB code\footnote{\tt http://camb.info/}
\cite{Lewis:1999bs}, 
based on CMBFast\footnote{\tt http://cfa-www.harvard.edu/$\sim$mzaldarr/CMBFAST/cmbfast.html}
\cite{Seljak:1996is} routines. We use 
CosmoMC\footnote{\tt http://cosmologist.info/cosmomc/} \cite{Lewis:2002ah} 
in order to sample the parameter space of our model  
with a Markov Chain Monte Carlo (MCMC) technique.

We assume a flat universe, with a constant equation
of state dark energy fluid, 
cold dark matter, 2
species of massless neutrinos plus a massive one, and ten free
parameters. Six of them are the standard $\Lambda$CDM
parameters, namely, the physical baryon density $\Omega_{b0} h^2$,
the physical cold dark matter density $\Omega_{c0} h^2 $, the
dimensionless Hubble constant $h$, the optical depth to
reionization $\tau_{\rm reion}$, the amplitude ($A_s$) and    
spectral index ($n_s$) of primordial density fluctuations.  In    
addition, we vary the constant dark energy equation of state    
parameter $w_{\phi}$ and the three parameters accounting for the  
neutrino mass: the present mass $m_0$,  
the logarithm of the parameter $\de$ related to the duration of
the transition, 
and the logarithm  of the ratio
of the modulus of the mass difference over the current mass, $\log\mu$,
where we define
\begin{equation*}
\ \mu \equiv \frac{|m_1 - m_0|}{m_0}  \!  \left\{ \begin{array}{llr}
\ \mu_{+} \equiv \frac{m_1}{m_0} - 1                    
&\! , \!&   m_1  >  m_0 \ ,\\[2mm]
\ \mu_{-} \equiv 1- \frac{m_1}{m_0}                    
&\! , \!&   m_1  <  m_0 \ .
\end {array} \right.
\end{equation*}
All these parameters take implicit flat priors in
the regions in which they are allowed to vary (see Table
\ref{tab:priors}).

%
\begin{table}[t]
\caption{Assumed ranges for the MaVaNs parameters \label{tab:priors} }
\begin{ruledtabular}
\begin{tabular}{cc}
Parameter & Range \\
\hline
$w_{\phi}$ & $-1<w_\phi<-0.5$ \\
$m_0$ &  $0<m_0/\eV<5$\\
$\de$ & $-4< \log\de<2$\\
$\mu$ & $-6< \log(\mu_{+})<0$ \\
 $\ $ & $-6< \log(\mu_{-})<0$
\end{tabular}
\end{ruledtabular}
\end{table}

%
\begin{table*}[]
\caption{ Results for increasing and decreasing neutrino mass, using WMAP 5yr $+$ small scale CMB $+$ LSS $+$ SN $+$ HST data. \label{tab:results}}
\begin{ruledtabular}
\begin{tabular}{ccc}
$\ $ & $ (+)$Region  95$\%$ (68$\%$) C.L. & 
$(-)$Region  95$\%$ (68$\%$) C.L.\\
\hline
$w_{\phi}$ &  $ < - 0.85$ \ ($< -0.91$) & 
$< - 0.87$ \ ($< -0.93$) \\
$m_0$ (eV) & $< 0.28$ \ ($< 0.10$) &  $ < 0.43$ \ ($< 0.21$) \\
$\log{\mu_{+}}$  &  $< -2.7 \ (< -4.5)$  &  \textemdash\\
$\log{\mu_{-}}$  & \textemdash  & $< -1.3 \ (< -3.1)$ \\
$\log\de$ &  $ [-3.84; 0.53] \ \ ([-2.20; 0.05])$ & $[-0.13; 4] \ \ ([0.56; 4])$ 
\end{tabular}
\end{ruledtabular}
\end{table*}

Concerning the last parameter, notice that we choose
to divide the parameter space between two regions: one
in which the mass is decreasing over time ($\mu_{+}$) and
one in which it is increasing ($\mu_{-}$). We chose to make this
separation because the impact on cosmological
observables is different in each regime, as we will discuss
later, and by analyzing this regions 
separately we can gain a better insight of the physics
driving the constraints in each one of them.  Moreover,
we do not allow for models with $w_\phi<-1$, since we are
only considering scalar field models with standard kinetic
terms. 

For given values of all these parameters, our modified version of
  CAMB first integrates the background equations backward in time, in
  order to find the initial value of $\rho_\phi$ leading to the
  correct dark energy density today. This problem does not always
  admit a solution leading to well-behaved perturbations: the dark
  energy perturbation equations (\ref{eq:depert1}), (\ref{eq:depert2})
  become singular whenever one of the two quantities, $\rho_\phi$ or
  $[1+\beta \rho_\nu (1-3 w_\nu)]$, appearing in the denominators
  vanishes. As we shall see later, in the case in which the neutrino
  mass decreases, the background evolution is compatible with cases in
  which the dark energy density crosses zero, while the second term
  can never vanish.  We exclude singular models by stopping the
  execution of CAMB whenever $\rho_\phi<0$, and giving a negligible
  probability to these models in CosmoMC. The physical interpretation
  of these pathological models will be explained in the next sections.
  For other models, CAMB integrates the full perturbation equations,
  and passes the CMB and matter power spectra to CosmoMC for
  comparison with the data. 

We constrain this scenario using CMB data (from WMAP
5yr~\cite{Komatsu:2008hk,Dunkley:2008ie}, VSA~\cite{Scott:2002th},
CBI~\cite{Pearson:2002tr} and ACBAR~\cite{Kuo:2002ua}); matter power
spectrum from large scale structure (LSS) data
(2dFGRS~\cite{Cole:2005sx} and SDSS~\cite{Tegmark:2006az}); supernovae Ia
(SN) data from ~\cite{Union2008}, and the HST Key
project measurements of the Hubble constant
\cite{freedman01}\footnote{While this work was being finished, 
the SHOES (Supernova, HO, for the Equation of State) 
Team \cite{Riess:2009pu} reduced the uncertainty on the Hubble constant by more than a factor 2
with respect to the value obtained by the HST Key Project, 
finding $H_0=74.2 \pm 3.6$ km s$^{-1}$ Mpc$^{-1}$. However, since we are taking a flat prior on $H_0$, and our best fit value
for $H_0$ is contained in their 1$\sigma$ region, we do not expect
our results to be strongly affected by their results.}.

\begin{figure}[b]
\begin{center}
\includegraphics[scale=0.65]{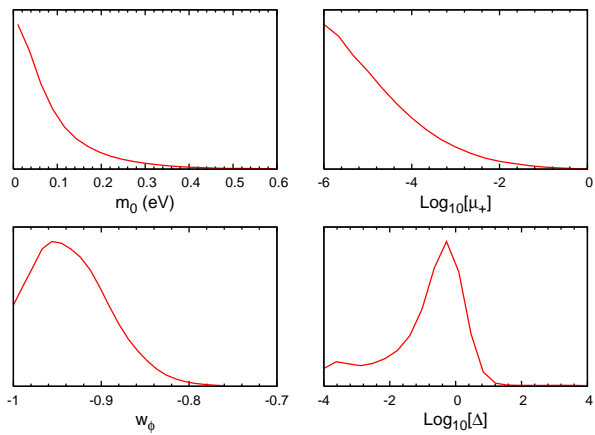} \\
\caption{ \small \label{fig:1Dplus} (Color online:) Marginalised 1D
  probability distribution (red/solid lines) for the decreasing mass
  case $m_1 > m_0$, for neutrino / dark energy parameters: $m_0$,
  $\log[\mu_{+}]$ (top panels), $w_{\phi}$, and $\log \Delta$ (bottom
  panels).}
\end{center}
\end{figure}

\begin{figure*}
\subfigure{
\includegraphics[scale=0.50, angle=-90]{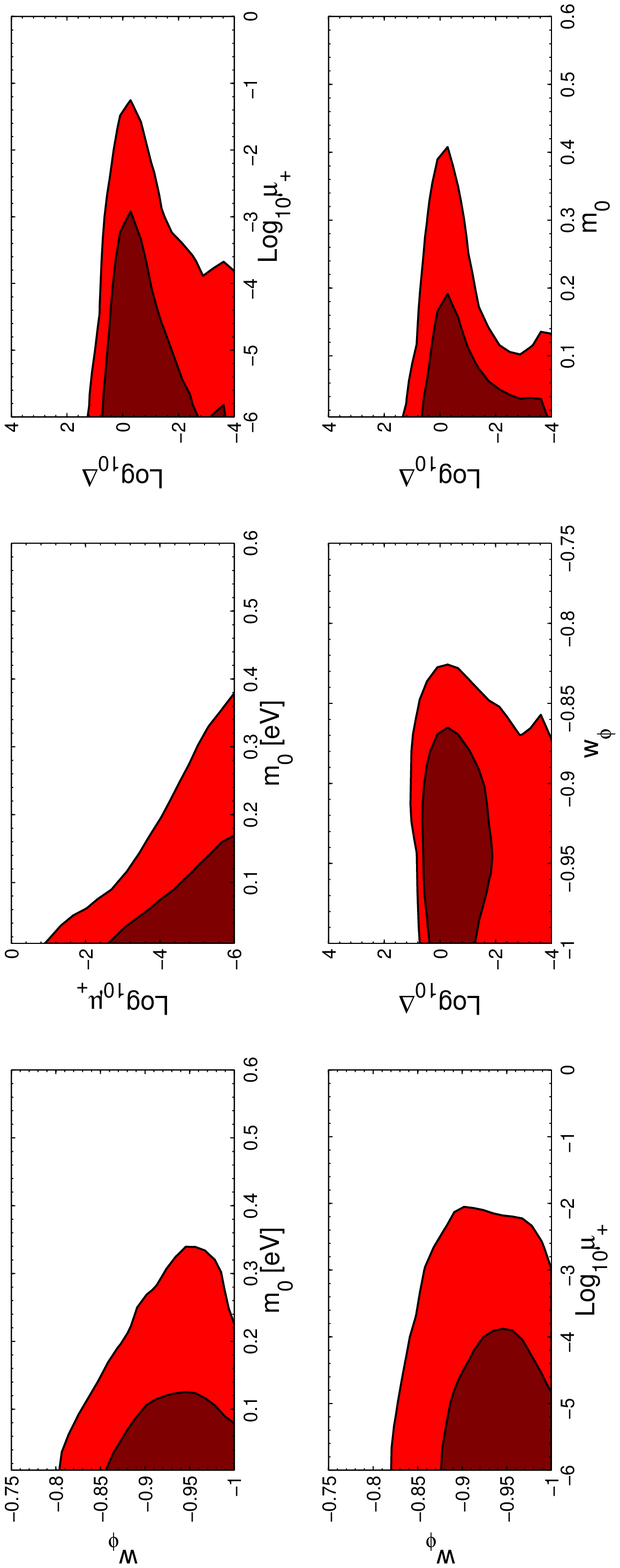}}
\caption{ \label{fig:2Dplus} (Color online) 
Marginalised 2D probability
distribution for decreasing mass, $m_1 > m_0$.
}
\end{figure*}

Once the  posterior probability of all ten parameters has
  been obtained, we can marginalize over all but one or two of them,
  to obtain one- or two-dimensional probability distributions. We
  verified that the confidence limits on the usual six
      parameters do not differ significantly from what is obtained in the
      ``vanilla model'' \cite{Komatsu:2008hk}, and therefore we only provide
      the results for the extra neutrino and dark energy parameters (Figures
\ref{fig:2Dplus},  \ref{fig:1Dplus}, \ref{fig:2Dminus}, \ref{fig:1Dminus}, and Table \ref{tab:results}).

\subsection{Increasing neutrino mass}

\begin{figure}[t]
\begin{center}
\includegraphics[scale=0.65]{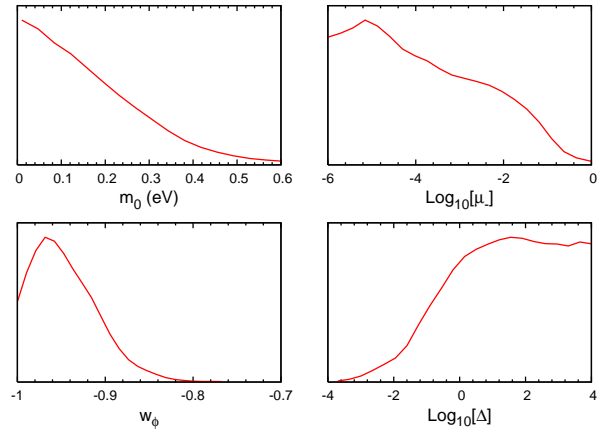} \\
\caption{ \small \label{fig:1Dminus} (Color online:) Marginalised 1D
  probability distribution in the increasing mass case $m_1 < m_0$,
  for the neutrino / dark energy parameters: $m_0$,
  $\log_{10}[\mu_{-}]$ (top panels), $w_{\phi}$, and $\log \Delta$
  (bottom panels).  }
\end{center}
\end{figure}

\begin{figure*}[t]
\subfigure{
\includegraphics[scale=0.50, angle=-90]{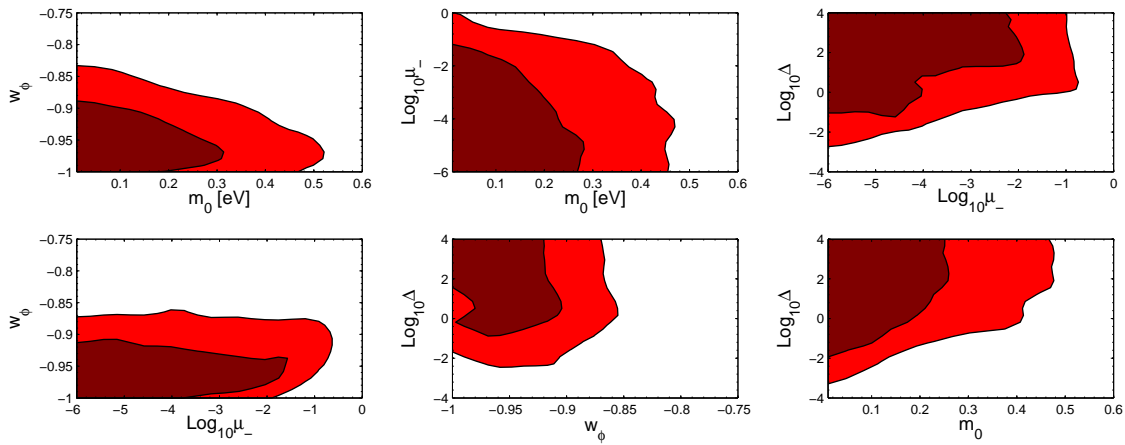}}
\caption{ \label{fig:2Dminus} (Color online) 
Marginalised 2D probability
distribution in the increasing mass case $m_1 < m_0$. 
}
\end{figure*}

In this model, the background evolution of the dark energy component obeys to 
equation (\ref{eq:deenergy}), which reads after division by $\rho_\phi$:
\bea
\frac{\dot{\rho}_\phi}{\rho_\phi} &=& - 3 {\cal H} (1 + w_\phi) 
- \frac {d \ln m_\nu}{du} \, \frac{\rho_\nu}{\rho_\phi} \, {\cal H} (1 - 3 w_\nu) \label{eq:rate} \\
&\equiv& - \Gamma_d - \Gamma_i \nonumber
\eea
where the two positive quantities $\Gamma_d$ and $\Gamma_i$ represent
respectively the dilution rate and interaction rate of the dark energy
density. For any parameter choice, $\rho_\phi$ can only decrease with
time, so that the integration of the dark energy background equation
backward in time always find well-behaved solutions with positive
values of $\rho_\phi$.  Moreover, the quantity $[1+\beta \rho_\nu (1-3
  w_\nu)]$ appearing in the denominator of the dark energy
perturbation equations is equal to the contribution of the dilution
rate to the total energy loss rate, $\Gamma_d / (\Gamma_d +
\Gamma_i)$. This quantity is by construction greater than zero, and
the dark energy equations cannot become singular.  However, when the
the interaction rate becomes very large with respect to the dilution
rate, this denominator can become arbitrarily close to zero.  Then,
the dark energy perturbations can be enhanced considerably, distorting
the observable spectra and conflicting the data. Actually, this
amplification mechanism is well-known and was studied by various
authors~\cite{Bean:2007ny,Valiviita2008,Gavela2009}.  It was found to
affect the largest wavelengths first, and is usually refered as the
large scale instability of coupled dark energy models.  The condition
for avoiding this instability can be thought to be roughly of the form
\be \label{eq:condition_increasing}
\Gamma_i < A \Gamma_d \ ,
\ee 
where $A$ is some number depending on the cosmological parameters and
on the data set (since a given data set tells how constrained is the
large scale instability, i.e. how small can be the denominator
$[1+\beta \rho_\nu (1-3 w_\nu)]$, i.e. how small should the
interaction rate remain with respect to the dilution rate).  The
perturbations are amplified when the denominator is much smaller than
one, so $A$ should be a number much greater than one.  Intuitively, the
condition (\ref{eq:condition_decreasing}) will lead to the rejection
of models with small values of ($w_\phi$, $\Delta$) and large values
of $\mu_-$. Indeed, the interaction rate is too large when the mass
variation is significant (large $\mu_-$) and rapid (small
$\Delta$). The dilution rate is too small when $w_\phi$ is small
(close to the cosmological constant limit). Because of that, it seems
that when the dark energy equation of state is allowed to vary one can 
obtain a larger number of viable models if $w_{\phi} > -0.8$ early on in the
cosmological evolution \cite{Majerotto:2009np,Valiviita:2009nu}.

We ran CosmoMC with our full data set in order to see how much this
mass-varying scenario can depart from a standard cosmological model
with a fixed dark energy equation of state and massive neutrinos. In
our parameter basis, this standard model corresponds to the limit
${\log} \mu_- \rightarrow - \infty$, with whatever value of
${\log} \Delta$. The observational signature of a neutrino
mass variation during dark energy or matter domination is encoded in
well-known effects, such as: (i) a modification of the small-scale
matter power spectrum [due to a different free-streaming history], or
(ii) a change in the time of matter/radiation equality [due to a
  different correspondence between the values of ($\omega_b$,
  $\omega_m$, $\omega_\nu$) today and the actual matter density at the
  time of equality]. On top of that, the neutrino and dark energy
perturbations can approach the regime of large-scale instability
discussed above.

\begin{figure*}[t]
\begin{center}
\includegraphics[scale=0.42]{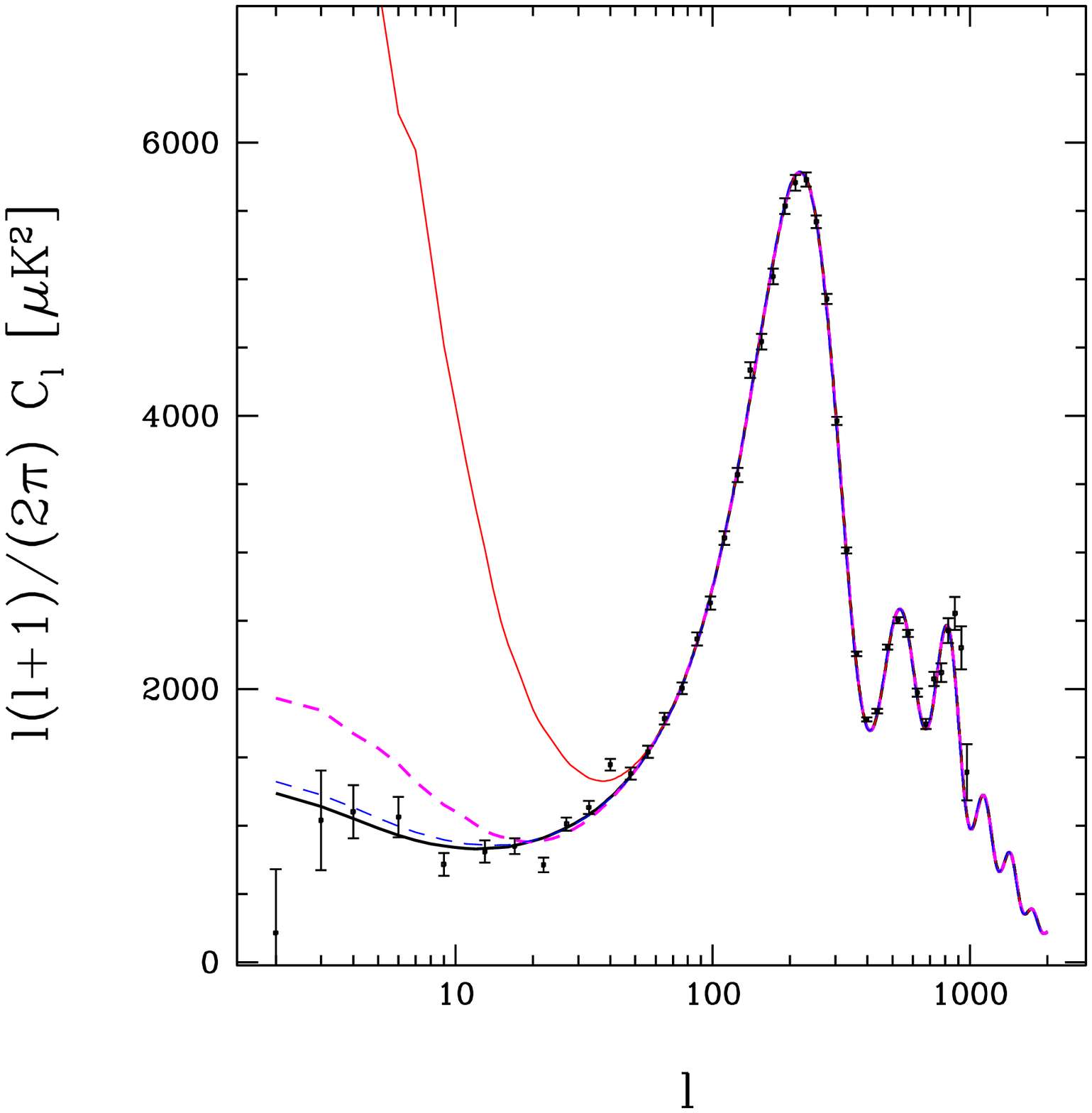} 
\includegraphics[scale=0.42]{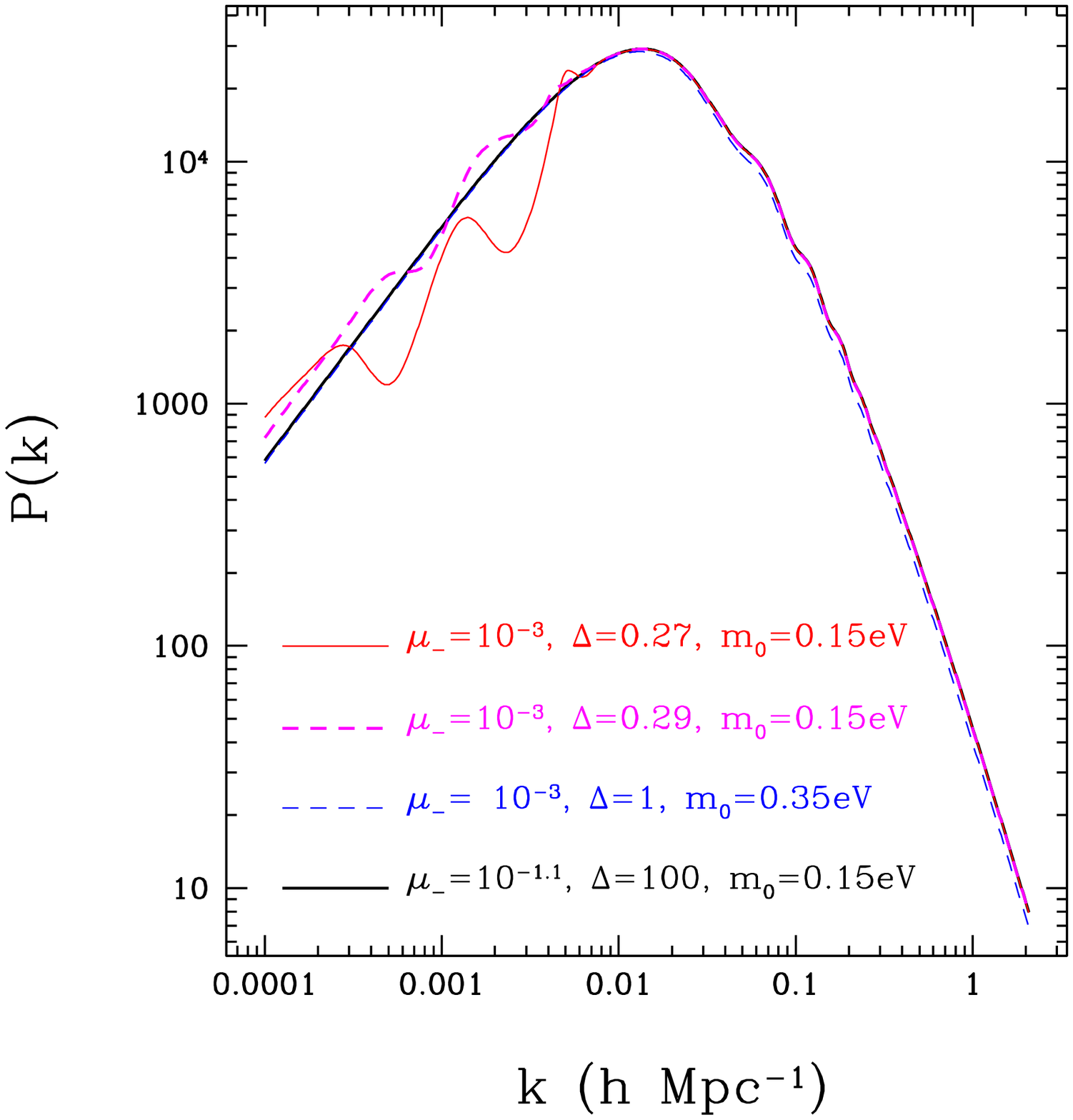} 
\caption{ \small  \label{fig:lsipower} 
(Color on-line)  CMB anisotropies and matter power spectra for some 
mass varying models with increasing mass, showing the development of the large scale instability. The 
cosmological parameters are set to our best fit values, except for the ones 
shown in the plot. The data points in the CMB spectrum correspond to the 
binned WMAP 5yr data. 
}
\end{center}
\end{figure*}

Our final results - namely, the marginalized 1D and 2D parameter
probabilities - are shown in figures \ref{fig:1Dminus} and
\ref{fig:2Dminus}.  The shape of the contours in $({\log} \mu_-, \log \Delta)$ space is easily
understandable with analytic approximations. The necessary condition
(\ref{eq:condition_increasing}) for avoiding the large-scale
instability reads in terms of our model parameters
\be \label{eq:lsi}
\mu_{-} \left[ \frac{1+ \Delta(1+\Gamma)}{\Delta}\right] < A
\left[\frac{1}{\left( 1- \Gamma \right) \left( 1 - f \right)}\right] 
\frac{3\Om_{\phi}(1+w_{\phi})}{\Om_{\nu}(1 - 3 w_{\nu})} \ ,
\ee
where we expressed the mass variation as
\be
\frac{d\ln m_{\nu}}{du} = \left(\frac{\mu_{-}}{1-\mu_{-}\Gamma}\right) \left( \frac{1+ \Delta}{\Delta}\right)
\left( 1- \Gamma \right) \left( 1 - f \right) ~.
\ee
Two limits can be clearly seen from this equations. For $\Delta \ll 1$
(fast transitions), the upper limit on $\mu_{-}$ reads
\be
\mu_{-} \lesssim A
\Delta \left[\frac{1}{\left( 1- \Gamma \right) \left( 1 - f \right)}\right] 
\frac{3\Om_{\phi}(1+w_{\phi})}{\Om_{\nu}(1 - 3 w_{\nu})} \ .
\label{eq:cond1}
\ee
This corresponds to the diagonal limit in the lower half of
the right upper panel of figure \ref{fig:2Dminus}. In fact, the
appearance of the large-scale instability is seen in models localized at the edge
of the allowed region, as shown in figure \ref{fig:lsipower}.

In the opposite case of a very slow transition, $\Delta \gg 1$, it is
clear from eq. (\ref{eq:lsi}) that the limit on $\mu_{-}$ should be
independent on $\Delta$,
\be
\mu_{-} \lesssim A
\left[\frac{1}{\left( 1- \Gamma \right) \left( 1 - f \right)}\right] 
\frac{3\Om_{\phi}(1+w_{\phi})}{\Om_{\nu}(1 - 3 w_{\nu})} \ .
\label{eq:cond2}
\ee
This limit corresponds to the almost vertical cut in the upper part of
the plane $(\log \mu_{-}, \log \Delta )$ (upper right panel,
fig. \ref{fig:2Dminus}).

These conditions are easier to satisfy when at the time of the
transition, $\Omega_\phi (1 + w_\phi)$ is large. So, in order to avoid
the instability, large values of $w_\phi$ are preferred.  However, it
is well-known that cosmological observables (luminosity distance
relation, CMB and LSS power spectra) better fit the data for $w$ close
to $-1$ (cosmological constant limit). In the present model, the role
of the large-scale instability is to push the best-fit value from -1
to -0.96, but $w_\phi=-1$ is still allowed at the 68\% C.L.

The main result of this section is that the variation of the neutrino
mass is bounded to be small, not so much because of the constraining
power of large-scale structure observations in the regime where
neutrino free-streaming is important (i.e., small scales), but by CMB
and LSS data on the largest scales, which provide limits on the
possible instability in DE and neutrino perturbations.

Indeed, for the
allowed models, the mass variation could be at most of order 10$\%$
for masses around $0.05$ eV, and less than 1$\%$ for masses larger
than $0.3$ eV: this is undetectable with small scale clustering data, showing that
the limit really comes from large scales.

With those results, we conclude that there is no evidence for a
neutrino mass variation coming from the present data. In fact, as for
most cosmological data analyses, the concordance $\Lambda$CDM model
remains one of the best fits to the data, lying within the 68$\%$ interval
of this analysis.

Nonetheless, better constraints will possibly be obtained with
forthcoming data, especially the ones that probe patches of the
cosmological ``desert'' between $z \simeq 1100$ and $z \simeq 1$, like
CMB weak lensing \cite{Lesgourgues:2005yv}, and/or cross-correlations
of different pieces of data, like CMB and galaxy-density maps
\cite{Lesgourgues:2007ix}.  We can estimate, for instance, what is the
favored redshift range for the neutrino mass variation according to
our results. Taking $m_0 = 0.1$ eV and the mean likelihood values for
$\log \de$ and $\log[m_1/m_0]$, one can see that the bulk of the mass
variation takes place around $z\sim 20$, a redshift that possibly will
be probed by future tomographic probes like weak lensing
\cite{Hannestad:2006as,Kitching:2008dp} and especially 21 cm
absorption lines
\cite{Loeb:2003ya,Loeb:2008hg,Mao:2008ug,Pritchard:2008wy}. Those will
help not only to disentangle some degeneracies in the parameter space,
but will also allow for direct probes of the neutrino mass in
different redshift slices.

\subsection{Decreasing neutrino mass}

In this case, the evolution rate of the dark energy density is still
given by equation (\ref{eq:rate}) but with an opposite sign for the
interaction rate: in can be summarized as 
\be
\frac{\dot{\rho}_\phi}{\rho_\phi} = - \Gamma_d + \Gamma_i~, 
\ee 
with $\Gamma_d$ and $\Gamma_i$ both positive.  In principle, the
interaction rate could overcome the dilution rate, leading to an
increase of $\rho_\phi$. Hence, the integration of the dark energy
evolution equation backward in time can lead to negative values of
$\rho_\phi$, and the prior $\rho_\phi>0$ implemented in our CAMB
version is relevant. Still, the denominator $[1+\beta \rho_\nu (1-3
  w_\nu)]$ can never vanish since it is equal to $\Gamma_d / (\Gamma_d
- \Gamma_i)$.

Well before before the transition, the interaction rate is negligible
and $\dot{\rho}_\phi$ is always negative. We conclude that $\beta = d
\ln m_\nu / d \rho_\phi$ starts from small positive values and increases. If 
the
condition
\be
\Gamma_i < \Gamma_d \label{eq:condition_decreasing}
\ee
is violated during the transition, $\dot{\rho}_\phi$ will cross zero
and become positive. This corresponds to $\beta$ growing from zero
to $+\infty$, and from $-\infty$ to some finite negative value. After
$\Gamma_i/\Gamma_d$ has reached its maximum, $\beta$ undergoes the
opposite evolution. Reaching $\rho_\phi=0$ is only possible if
$\rho_\phi$ has a non-monotonic evolution, i.e. if
(\ref{eq:condition_decreasing}) is violated. However, the
perturbations diverge even before reaching this singular point: when
$\beta$ tends to infinity, it is clear from eq. (\ref{eq:hierarchy})
that the neutrino perturbation derivatives become arbitrarily
large. We conclude that in this model, the condition
(\ref{eq:condition_decreasing}) is a necessary condition for avoiding
instabilities, but not a sufficient condition: the data is expected to
put a limit on the largest possible value of $\beta$, which will
always be reached before $\dot{\rho}_\phi$ changes sign, i.e. before
the inequality (\ref{eq:condition_decreasing}) is saturated.  Hence,
the condition for avoiding the instability is intuitively of the form
of (\ref{eq:condition_increasing}), but now with $A$ being a number
smaller than one.

We then ran CosmoMC with the full data set and obtained the
marginalized 1D and 2D parameter probabilities shown in figures
\ref{fig:1Dplus} and \ref{fig:2Dplus}. The major differences with
respect to the increasing mass case are: a stronger bound on $m_0$, a
much stronger bound on $\mu_-$, and the fact that large values of
$\Delta$ are now excluded. This can be understood as follows.  In
order to avoid instabilites, it is necessary to satisfy the
inequalities (\ref{eq:cond1}), (\ref{eq:cond2}), but with a much
smaller value of $A$ than in the increasing mass case; hence, the
contours should look qualitatively similar to those obtained
previously, but with stronger bounds. This turns out to be the case,
although in addition, large $\Delta$ values are now
excluded. Looking at the mass variation for large $\Delta$ in figure
\ref{fig:mvn}, we see that in this limit the energy transfer takes
place essentially at low redhsift. Hence, the interaction rate is
large close to $z=0$.  In many models, this leads to positive
values of $\dot{\rho}_\phi$ at the present time, to a
non-monotonic behavior of the dark energy density, and to diverging
perturbations. This can only be avoided when $w$ is large with respect
to -1, i.e. when the dilution rate is enhanced. Hence, in this model,
the need to avoid diverging perturbations imposes a strong parameter
correlation between $w$ and $\Delta$. However, values of $w$ greater
than -0.8 are not compatible with the supernovae, CMB and LSS data set;
this slices out all models with large $\Delta$.

The fact that the bound on $m_0$ is stronger in the decreasing mass
case is also easily understandable: for the same value of the mass
difference $\mu_\pm=|m_1-m_0|/m_0$, a given $m_0$ corresponds to a larger
mass $m_1$ in the decreasing mass case. It is well-known that CMB and
LSS data constrain the neutrinos mass through its background effect,
i.e. through its impact on the time of matter/radiation equality for a
given dark matter abundance today. The impact is greater when $m_1$ is
larger, {i.e.} in the decreasing mass case; therefore, the bounds on $m_0$
are stronger.

\section{Concluding remarks} \label{sec:conc}

In this work we analysed some mass-varying neutrino scenarios in a
nearly model independent way, using a general and
well-behaved parameterization for the neutrino mass,
including variations in the dark energy density in a
self-consistent way, and taking neutrino/dark energy perturbations
into account. 

Our results for the background, CMB anisotropies, 
and matter power spectra are in agreement with previous analyses of 
particular scalar field models, showing that
the results obtained with this parameterization are robust and 
encompass the main features of the MaVaNs scenario.

Moreover, a comparison with cosmological data shows that only small
mass variations are allowed, and that MaVaNs scenario are mildly
disfavored with respect to the constant mass case, especially when
neutrinos become lighter as the universe expands.  In both cases,
  neutrinos can change significantly the evolution of the dark energy
  density, leading to instabilities in the dark energy and/or neutrino
  perturbations when the transfer of energy between the two components
  per unit of time is too large. These instabilities can only be
  avoided when the mass varies by a very small amount, especially in
  the case of a decreasing neutrino mass. Even in the case of
  increasing mass, constraining better the model with forthcoming data
  will be a difficult task, since it mimics a massless neutrino
scenario for most of the cosmological time. 

One should keep in mind that our analysis assumes a constant equation of
state for dark energy and a monotonic behavior
for the mass variation.  Even though those
features are present in most of the simplest possible models, more complicated models
surely can evade the constraints we obtained in our analysis.

Finally, those constraints will improve with 
forthcoming tomographic data. If any of the future probes indicate a mismatch in 
the values of the neutrino mass at different redshifts, we
could arguably have a case made for the mass-varying models.

\section*{Acknowledgments}
We would like to thank Luca Amendola, Alberto Fern\'{a}ndez-Soto,
Gennaro Miele, Miguel Quartin, Rogerio Rosenfeld, and Jos\'e W.F.\ Valle for
discussions concerning an earlier version of this work.  This work was supported
by the European Union (contracts No.\ RII3-CT-2004-506222 and
MRTN-CT-2004-503369, Marie Curie Training Network ``UniversetNet''
MRTN-CT-2006-035863), by the Spanish grants FPA2008-00319 (MEC) and
PROMETEO/2009/091 (Generalitat Valenciana), and by a MEC-IN2P3 agreement. 
UF is supported by an I3P-CSIC
fellowship. This work made some progress during a fruitful stay at
the Galileo Galilei Institute for Theoretical Physics, supported by
INFN. We also acknowledge the use of the Legacy Archive for Microwave
Background Data Analysis (LAMBDA). Support for LAMBDA is provided by
the NASA Office of Space Science.


\end{document}